\documentstyle[onecolumn]{mn}           
\input{psfig}

\title [What is a Peculiar Galaxy]{WHAT IS A PECULIAR GALAXY ?}

\author [Naim and Lahav]{A. Naim$^{1}$ \& O. Lahav$^{2}$\\ \\
$^1$ Dept. of Physics and Astronomy, The Johns Hopkins University, \\
Baltimore, MD, 21218 U.S.A.\\
$^2$ Institute of Astronomy, Madingley Rd., Cambridge, CB3 0HA, U.K. \\}

\date{Submitted to MNRAS ?? ????? 1996 - Revised ?? ????? 1996}

\begin{document}

\maketitle

\begin{abstract}

Following the recent surge of interest in peculiar galaxies at high redshifts 
we consider the definition, or lack thereof, of morphological peculiarities
on a sample of local universe galaxies. Studying the morphology of local 
universe galaxies is also of interest in trying to understand galaxy dynamics 
and quantifying the relations between morphology and environment. We use 
classifications given by five experts for a sample of 827 APM galaxies and find
that there is little agreement between them on what qualifies as a peculiar 
galaxy. We attempt several objective approaches : matching galaxy images to 
``templates''; examinig the $180^{\circ}$ Asymmetry against Light Concentration
(following Abraham {\it et al.} 1995); and exploring angle-dependent asymmetry 
measures. While none of the quantities we use results in a clean distinction 
between normal and peculiar galaxies, there is a rough correlation between some
parameters and image peculiarity. However, the mixing between the two classes 
is significant. We conclude that the class of peculiar galaxies is not totally 
distinct from the class of normal galaxies, and that what we are seeing is 
really a sequence.  It is therefore more useful to consider distribution 
functions of morphological parameters. The current and possibly other, more 
accurate parametrisations require better data, which is becoming available 
through CCD imaging. 

\end{abstract}

\section{Introduction}

Most of the observed galaxies in the local universe appear to form a 
morphological sequence, known as the Hubble sequence (e.g., Sandage 1961; de
Vaucouleurs {\it et al.} 1959). The defining criteria for various types of 
galaxy along the Hubble sequence are mostly qualitative rather than 
quantitative, and thus somewhat vague (e.g., Mihalas \& Binney, 1981). 
Nevertheless, in a recent study (Lahav {\it et al.} 1995; Naim {\it et al.} 
1995a) six experts were asked to classify the same sample of 835 galaxies from 
the APM Equatorial Catalogue (Raychaudhury {\it et al.}, in preparation) and 
came to a reasonable agreement with each other~: The rms dispersion over all 
pairs of experts was 1.8 types, on a scale of 16 types ranging from -5 for
Ellipticals to +10 for Irregulars. The notion of a well established 
morphological sequence pertains to what are loosely referred to as ``normal 
galaxies''. In addition to these, a certain (small) fraction of local universe 
galaxies are regarded ``peculiar'' to varying extents. In an attempt to better
understand spiral arms, H. Arp has collected into what became his Atlas of 
Peculiar Galaxies (1966) many examples of galaxies which deviated in some way
from the normal Hubble sequence. The morphological diversity of these peculiar
galaxies is quite impressive.

So long as the fraction of peculiar galaxies was considered very small, they
could be treated as individual cases that may not fit comfortably on the 
Hubble Sequence. This situation is now changing with the advent of imaging at
higher redshifts with the Hubble Space Telescope and large ground based 
telescopes. As has recently been reported (e.g., Griffiths {\it et al} 1994; 
Driver {\it et al} 1995; Abraham {\it et al} 1995), the fraction of peculiar
and irregular galaxies at moderate redshifts appears to be much higher than in 
the local universe. However, the numbers quoted rely on eyeball classification 
by different observers, and since the notion of peculiar galaxies is even more 
fuzzy than that of morphologically ``normal'' galaxies, any reference to the 
fraction of peculiar galaxies (at any redshift) may be very sensitive to the
adopted definition of peculiarity. Whatever definition is adopted, it is
crucial that a set of {\it objective} parameters be found that describe galaxy
morphologies faithfully at various redshifts. Automating the process of 
morphological analysis is inevitable in view of the very large numbers of
galaxy images now available, and an objective parametrisation is the only way
to do it.

In this paper we attempt to answer three fundamental questions :
\begin{enumerate}
\item{What is a peculiar galaxy ? What definition exists and should it be 
adopted or modified ?}
\item{Do peculiar galaxies constitute a separate class of galaxies, or do they
form a continuum that blends into the realm of normal galaxies ?}
\item{Which {\it objectively measured} parameters should one use in order to
define distribution functions of galaxy morphologies at all redshifts and for 
all inclinations ?}
\end{enumerate}

We treat the first question in \S~2, where we describe and compare the experts'
classifications. We treat the other two questions in \S~3, in which we discuss 
ways to quantify peculiarity and describe our results using several 
parametrisations. The discussion is brought in \S~4.

\section{Peculiarities from Eyeball Classifications}

When dealing with eyeball classifications of peculiar galaxies it is important 
to note that the term ``peculiar'' refers to a different kind of morphology 
than the term ``irregular'', although there appears to be some confusion 
between the two in the literature. Irregular galaxies lie at the late type end 
of the Hubble sequence. They are rather faint and fuzzy, and have little or no 
bulges. Peculiar galaxies are all those galaxies whose morphology is not 
similar to that of any of the Hubble types, including the irregulars. Our 
attempts are concentrated on telling peculiars apart from normal galaxies
and we include irregulars in the latter category. 

From an inspection of Arp's atlas (1966) it appears that in very broad terms
there are two kinds of morphological peculiarities : What we will refer to as
"mild" peculiarities are deviations from the "normal" patterns expected from 
a galaxy of any given Hubble type, which nevertheless allow the classification 
of that galaxy along the Hubble sequence. Examples include dust lanes in 
Ellipticals and three-armed Spirals. What we refer to as "strong" peculiarities
are severe disruptions of the general appearance of a galaxy, to such an extent
that normal classification is either very inconclusive or totally impossible. 
A galaxy with a mild peculiarity may be assigned a Hubble type with the
addition of a lower case ``p'' (e.g., SBcp). For strong peculiarities the type 
is simply ``Pec''. However, in most cases eyeball classifications are given
without explicitly defining the criteria, and this leads to confusion between
peculiars and normals as well as between different extents of peculiarity.

We seek to put the notions of mild and strong peculiarities on more solid,
quantitative grounds. As a first step we examine 827 galaxies from the sample 
of APM galaxies which was used in the comparative study of morphological 
classifications (Naim {\it et al.} 1995a). Five of the participating experts 
(R. Buta, H. Corwin, G. de Vaucouleurs, J. Huchra and S. van den Bergh) gave a 
full morphological description in addition to the numerical T-type. Of the 827 
images, 222 (27 \%) were flagged with some kind of a peculiarity by at least 
one expert. Only 22 galaxies (3 \%) were flagged as strongly peculiar by at 
least one expert, which indeed shows that strongly peculiar galaxies are quite 
rare in the local universe. However, the agreement between the experts on the 
issue of peculiarity is not very high : of the 222 galaxies flagged for some 
kind of peculiarity, 125 were flagged by only one expert, 44 by two, and only 
53 by more than two experts. {\it In only five cases did all the experts agree 
that the galaxy was peculiar}. The situation is not different when one 
restricts oneself to galaxies flagged as strongly peculiar by at least one 
expert~: of the total of 22 such galaxies, 8 were flagged as ``Pec'' by two 
experts and only two by three experts. There were no cases of four or five 
experts agreeing on a strong peculiarity. These low rates of agreement between 
experts mean that {\it there is no accepted definition for morphological 
peculiarities}. 

\section{Quantitative Measures  of Peculiarities}

Eyeball classifications of peculiarities appear to yield answers that are 
observer-dependent, and in addition are not easily applicable to large numbers 
of images. As suggested in our previous work on morphological classification 
of normal galaxies (Lahav {\it et al.} 1995; Naim {\it et al.} 1995b), 
automating the process is essential. Here we focus on quantitative parameters 
for the description of peculiarities.

Identifying a mild peculiarity could be approached by matching an image with a 
set of templates, each describing a certain morphological type. Work along 
these lines has been done (e.g., Spiekermann 1992) for the classification of 
normal galaxies. However, this raises an even more difficult question, namely, 
what is a {\it normal} galaxy? In other words, what variety of galaxy 
templates, and in particular how many of them, can be considered as sufficient 
to represent all normal galaxies as distinct from peculiars ? we try to answer 
this question below.

On the other hand, detecting a strong peculiarity may be an easier exercise,
but nevertheless requires a quantitative definition. Abraham et al. (1995)
claim that peculiar (and irregular) galaxies are much less symmetric 
than normal galaxies, and suggest an asymmetry parameter ($A$) which is based 
on rotating the image by $180^\circ$ and subtracting the result from the 
original image. This single parameter was augmented by the central 
concentration of light ($C$) parameter, which was originally suggested by 
Morgan (1958) and later used successfully by Kodaira {\it et al.} (1986). In 
effect, using the light concentration as a key parameter implies the a-priori 
assumption that peculiarities are related to late-type, disk-dominated galaxies
only. We therefore examine the $A$ parameter alone, apply it to the APM sample 
and generalise it into a set of five parameters. 

\subsection{Identifying Peculiarities by Template Matching}

As a first guess at the number of galaxy templates required in order to span
the full range of normal morphologies, we selected galaxies from the APM sample
according to two criteria : (i) They were given definite classifications by
at least five experts (As noted in Naim et al. (1995a), the experts reserved 
the right not to classify a galaxy at all, or give a partial classification 
such as "S", when they were very uncertain); (ii) The rms dispersion between 
the experts for every selected galaxy was less than 0.5 types. We then 
inspected the selected images by eye, and removed from the list cases of 
morphologies that were nearly identical to other members in the list. The goal 
was to end up with as diverse a collection of morphologies and inclinations as 
possible, while keeping the overall number of templates relatively small. The 
resulting list contained 35 galaxies, which represent the following seven 
classification bins (according to their mean types, given by the experts) : 

\smallskip
\begin {enumerate}
\item{Types [-5,-3.5] : E, 3 galaxies}
\item{Types (-3.5,0.0) : S0, 3 galaxies}
\item{Types [0.0,2.0) : Sa, 4 galaxies}
\item{Types [2.0,4.0) : Sb, 8 galaxies}
\item{Types [4.0,6.0) : Sc, 9 galaxies}
\item{Types [6.0,8.0) : Sd, 3 galaxies}
\item{Types [8.0,10.0] : Sm/Ir, 5 galaxies}
\end{enumerate}

Any measure of similarity between two images should be insensitive to size
and inclination differences. To ensure this, we used the image reduction 
software developed for the APM galaxies (Naim {\it et al.} 1995b). As a first
stage we detected foreground stars superimposed on the galaxy image and removed
them. Since the images were taken from plates, most stellar images were 
saturated and no point-spread-function could be well fitted to them. This
means that both the detection of stars and their removal were rather crude.
We then sampled each image on a series of $N_e = 30$ ellipses, all sharing the 
ellipticity and position angle of the whole image, but each with a different 
semi-major axis length. The difference $D_{ij}$ between any two images $i$ and 
$j$ was defined as :

\bigskip
\begin{math}
$$
(1)~~~~~~~~~~~~~ D_{ij}^2 = \frac{1}{N_e} \Sigma_{k=1}^{N_e} \frac
                       {\Sigma_{l=1}^{N_k} (I^{i}(k,l) - I^{j}(k,l))^2} {N_k}
$$
\end{math}

\medskip
\noindent where the outer sum is over the $N_e$ sampled ellipses, the inner sum
is over the $N_k$ points of ellipse $k$, and $I(k,l)$ denotes the sampled 
intensity at point $l$ of ellipse $k$.

Classification of a given galaxy image was then carried out by calculating the 
difference between that galaxy and every template galaxy. The difference was
also calculated with the template galaxy reflected about its minor axis, in
order to be reflection-independent, and smaller of the two numbers was taken as
the difference between the two images. Typically, our current image will 
resemble one template of a given type more than other templates of the same
type. For this reason the difference between our image and the entire type is
taken as the smallest of its differences from all the templates belonging to
that type. In this way we express the classification of any galaxy in the 
sample by a 7-dimensional vector of these minimal distances. For normal 
galaxies the smallest component of this vector ought to be the one 
corresponding to the correct classification bin, and peculiar galaxies are 
expected to have large distances from all bins. We define as strongly peculiar 
galaxies having at least one expert assign them a ``Pec'' type, while mildly 
peculiar galaxies are defined to have no ``Pec'' assignment and at least three 
``p'' assignments by the experts. With these definitions we have 34 mild 
peculiars and 22 strong peculiars in the sample.

Figure 1 depicts histograms of the minimal component of the distances vector 
for the three populations of galaxies. Naively, one would expect the minimal 
difference to be large for peculiars (not really close to any of the classes), 
while normal galaxies should be much closer to one bin, so the minimal value 
should be much smaller. However, as can be clearly seen from the figure, the 
peculiars largely follow the trend set by the normals, although some of the 
strong peculiars do exhibit the expected behaviour. The medians of each plot 
are indicated by a vertical dashed line, and do not change much from one plot 
to the other : The median of the normal galaxies is $0.078$, rising slightly to
$0.083$ for mild peculiars and up to $0.9$ for strong peculiars. It appears 
that our collection of 35 templates does not truly represent the morphological 
variety of the whole sample. We tried to solve this problem by increasing the 
number of templates we use. We repeated the above exercise for a collection of 
262 templates, which were selected by relaxing the largest allowable rms of the
experts' classifications from 0.5 to 0.8 types, but still requiring five 
definite classifications. The templates were divided among the seven 
classification bins as follows : 

\smallskip
\begin {enumerate}
\item{Types [-5,-3.5] : E, 11 galaxies}
\item{Types (-3.5,0.0) : S0, 19 galaxies}
\item{Types [0.0,2.0) : Sa, 29 galaxies}
\item{Types [2.0,4.0) : Sb, 98 galaxies}
\item{Types [4.0,6.0) : Sc, 81 galaxies}
\item{Types [6.0,8.0) : Sd, 11 galaxies}
\item{Types [8.0,10.0] : Sm/Ir, 13 galaxies}
\end{enumerate}

Figure 2 shows the results in this case. It appears that the use of more 
templates has made only a small difference : The medians are now at $0.044$ for
normals, $0.068$ for mild peculiars and $0.083$ for strong peculiars. While
the differences between the medians are larger than before, the mixture between
the populations is still large. It seems therefore that template matching does 
not work properly for the distinction between normal and peculiar galaxies,
although it is difficult to decouple it from the uncertainties in the experts' 
classifications.

\subsection {Suggested Measures of Peculiarity}

In an attempt to decouple the effect of vague classifications from that of 
inadequate parametrisation, we now turn to look for specific features which 
distinguish normal galaxies from peculiars. We examine two sets of parameters 
here : One was suggested by Abraham {\it et al.} (1994, 1995) and comprises 
the light concentration index and the asymmetry parameter (the C-A pair). We 
adopt the definitions they give for both parameters. The other set is a 
generalisation of their asymmetry parameter : We use the $N_e = 30$ sampled 
ellipses and define 32 radial rays, of constant angular separations ($\pi/16$).
Instead of looking at differences due to $180^{\circ}$ rotations only, we then 
examine five different angular separations : $\theta = 180^{\circ}, 90^{\circ},
45^{\circ}, 22.5^{\circ}, 11.25^{\circ}$. Figure 3 shows schematically all the
rays, a selected pair of which (at separation of $11.25^{\circ}$ are 
highlighted. For any two such radial rays, $k$ and $l$, we define the 
difference $d(k,l)$ as :

\bigskip
\begin{math}
$$
(2)~~~~~~~~~~~~~ d^2 (k,l) =  \frac {1}{N_e} 
                           \Sigma_{m=1}^{N_e} [I(k,m) - I(l,m)]^2
$$
\end{math}

\medskip
\noindent where $I(k,m)$ is the sampled point at the intersection of radial
ray $k$ and ellipse $m$. The asymmetry at angular separation $\theta$ is then 
given by :

\bigskip
\begin{math}
$$
(3)~~~~~~~~~~~~~ d_{\theta}^2 =  \frac {1}{N_{\theta}} \sigma_{(k,l) | 
                               \theta} ~d^2 (k,l)
$$
\end{math}

\medskip
\noindent where the summation is over all $N_{\theta}$ pairs of rays $(k,l)$ 
such that their angular separation is $\theta$. It is important to note that
the $A$ parameter is calculated on the entire image, while the generalised 
parameters are measured on rays selected from the sampled ellipses.

\subsection{Applying the Parameters to APM Images}

Figure 4 depicts the positions on the C-A plane occupied by normal, mildly
peculiar and strongly peculiar galaxies (defined as above). The dashed lines
in each plot mark the median values of each of the two parameters. The picture 
is essentially the same as before : there is a high degree of mixing between 
normals and peculiars, and telling the two kinds of peculiars apart is 
virtually impossible.

In figures 5-7 we show the results of applying the generalised asymmetry
parameters. Figure 5 shows histograms depicting asymmetry values for each of
the five angular separations, for normal galaxies. Figure 6 shows the 
corresponding histograms for mild peculiars, while strong peculiars are
depicted in figure 7. The distribution of normal galaxies is concentrated
around low values in all angular separations, as one would expect. The 
peculiars, in particular the strong peculiars, show an almost bimodal 
distribution, with the majority of galaxies overlapping on the positions held 
by normal galaxies and a smaller fraction lying at higher degrees of 
difference. The medians of these histograms are again shown as dashed vertical 
lines, and figure 8 shows the medians as a functions of angle of separations 
for all three populations. The gap between the medians is largest at angular 
separation of $45^{\circ}$, decreasing towards both ends.

It is very interesting to note that all five differences correlate with each 
other : Principal components analysis of these data shows that $91\%$ of the 
variance are spanned by the first PC. Galaxies appear to exhibit roughly the
same degree of asymmetry at all angular separations. However, these are
tentative conclusions drawn from survey plate material. It may well be that
repeating this exercise with CCD images will give a different conclusion.
In figure 9 we show scatter plots of the five differences against the $A$ 
parameter. The correlations are significant but the scatter is large. The 
$180^{\circ}$ difference appears to have a shallower slope than the others with
a hint of curvature, and its scatter is lower than that of the others. The
scatter increases as one goes to lower angular separations. The five 
generalised asymmetries therefore convey more information than just the $A$ 
parameter. Again, using CCD images these parameters may indicate more 
diversity.

Figures 10 and 11 show the strong and mild peculiars in our sample. Figure 12
shows a set of normal galaxies from the sample, for which a high $45^{\circ}$
difference ($> 0.2$) was derived. Some of the objects in figure 12 look similar
to some peculiars in the sample, while others appear normal but are 
contaminated by foreground stars. As mentioned above, the removal of stars was
crude and may have been insufficient in some cases (especially for edge-on
galaxy images).

\section{Discussion}

The nature of peculiar galaxies depends on their definition. In the absence of
a definition we look to expert classifiers for agreed examples. However, even
the experts do not agree very well on this issue. The conclusion is that the
so-called class of peculiar galaxies is in fact a mixed bag of deviations
from what are regarded as normal morphologies. In some cases peculiarity 
manifests itself through asymmetry, but in other cases it is much more subtle 
and difficult to detect. The transition between normal and peculiar galaxies 
appears to be very smooth. One can imagine the locus of normal galaxies in the 
centre of morphology space, with various deviations stretching in different 
directions. The further away from the centre one goes in a certain direction 
the more peculiar the corresponding morphology. Since deviations from 
"normality" form a continuum, it is down to the observer who classifies the 
galaxy to draw the line beyond which a galaxy is called peculiar. The answer 
to the question "what is a peculiar galaxy" is therefore not only a matter of 
defining what deviations from normality to look for, but also of setting the 
extent of the deviation. Alternatively, and more sensibly, one can do away
with classifications and simply look at distributions of galaxies in morphology
space.

The variety of galaxy morphologies is so great that qualifying peculiarity
through template matching can not succeed, unless one is willing to employ
a vast collection of templates that incorporates each and every "normal"
morphology. This is not practical, both technically (requiring too many
eyeball classifications) and conceptually (there is no clear quantitative
definition of normal galaxies, either). The alternatives tried by Abraham {\it 
et al} (1995) and by ourselves manage to capture only one kind of deviation 
from normality : a general distortion of the shape of the galaxy. Even then, 
"noisy" normal galaxies are mixed with truly distorted peculiars. Noise
sources include contamination by foreground stars which were not fully removed.
The inclination of a galaxy image also plays a role in determining how 
accurately parameters are measured. The problem of star removal could be dealt 
with on CCD images, where one can calculate a stellar point-spread function and
subtract it. The quality of data is therefore of great importance, and the 
increasing numbers of CCD-based surveys will help in avoiding problems such as 
insufficient dynamic range and non-linear response, which are typical of plate 
material. It is certainly worth while revisiting the analysis of this paper 
with CCD images of galaxies.
 
\bigskip

{\bf Acknowledgements}

We would like to thank the following people for constructive discussions and 
useful suggestions : Bob Abraham, Richard Ellis, ,Somak Raychaudhury, Laerte
Sodr{\'e} Jr. and Michael Storrie-Lombardi.

\newpage

\bigskip
Figure Captions :

\bigskip
Figure 1 : Minimal Difference vs. Classification Bin, using 35 templates, 
  for three kinds of galaxies. Top Left : Normals; Top Right : Mildly Peculiar;
  Bottom Right : Strongly Peculiar. 

\bigskip
Figure 2 : Minimal Difference vs. Classification Bin, using 262 templates, 
  for three kinds of galaxies. Top Left : Normals; Top Right : Mildly Peculiar;
  Bottom Right : Strongly Peculiar. 

\bigskip
Figure 3 : Schematic Description of the Radial Rays used in Calculating the
  Generalised Asymmetry Parameters.

\bigskip
Figure 4 : Distribution of Normal and Peculiar Galaxies on the C-A Plane. The
  medians of each parameter are also indicated.

\bigskip
Figure 5 : Distribution of the Generalised Asymmetry Parameters for Normal 
Galaxies.

\bigskip
Figure 6 : Distribution of the Generalised Asymmetry Parameters for Mildly 
Peculiar Galaxies.

\bigskip
Figure 7 : Distribution of the Generalised Asymmetry Parameters for Strongly 
Peculiar Galaxies.

\bigskip
Figure 8 : Medians of Generalised Asymmetry for Normal and Peculiar Galaxies.

\bigskip
Figure 9 : Generalised Asymmetry Parameters vs. the A Parameter.

\bigskip
Figure 10 : The Strong Peculiars in our sample.

\bigskip
Figure 11 : The Mild Peculiars in our sample.

\bigskip
Figure 12 : Asymmetric Normal Galaxies in the Sample.

\end{document}